\documentclass[aps,pra,twocolumn,groupedaddress,nofootinbib]{revtex4-2}

\usepackage{amsmath,amssymb,mathtools,bm}
\usepackage{amsthm}
\newtheorem{theorem}{Theorem}
\newtheorem{corollary}{Corollary}
\usepackage{booktabs}
\usepackage{graphicx}
\usepackage{microtype}
\usepackage{hyperref}
\usepackage{physics}

\usepackage[nolinenumbers,noshowlabels]{jadeleon} 
\DeclareCommenter{cp}{0000FF}{CP}

\hypersetup{
  colorlinks=true,
  linkcolor=blue,
  citecolor=blue,
  urlcolor=blue
}

\renewcommand{\Tr}{\operatorname{Tr}}
\newcommand{\supp}{\operatorname{supp}}

\newcommand{\id}{\mathbb{I}}
\newcommand{\C}{\mathbb{C}}

\newcommand{\corref}[1]{Corollary~\ref{#1}}

\begin{document}
\title{State convertibility in quantum-state texture from a general form of free operations}

\author{Jose Alfredo de Leon}
\affiliation{\ifunam}

\begin{abstract} 
State convertibility specifies which transformations between states a quantum
resource theory allows.
In quantum-state texture, deterministic conversion criteria are known for
qubits and pure sources in arbitrary dimension, but a general criterion
for mixed states in higher dimensions is missing.
We derive this exact criterion from a complete characterization of free
operations in arbitrary finite dimension.
Because the unique free state is pure, every free operation splits into
a quantum filter and a residual completely positive map. 
Using this decomposition, we show that every allowed conversion admits
a two-outcome measurement that retains the filtered state after one outcome
and prepares a fixed state after the other.
Conversion thus reduces to finding such a filter whose unnormalized output
leaves a positive semidefinite remainder when subtracted from the target, a search we show to be
solvable by semidefinite programming.
Beyond deterministic conversion, the characterization of free operations can be used
to study probabilistic and assisted manipulation of quantum-state texture.
More broadly, since the construction relies on a pure fixed state, it also applies to
other resource theories defined by free operations fixing such a state.
\end{abstract} 
\maketitle
\begin{figure*}[t] 
  \centering
  \includegraphics[width=\textwidth]{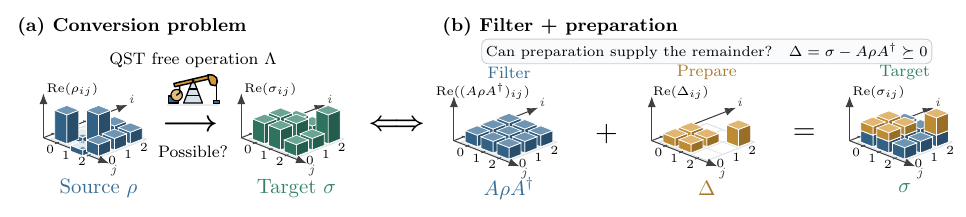}
  \caption{QST conversion problem and main results.
  (a) When can a source state $\rho$ be converted into a target $\sigma$ by a QST free
  operation $\Lambda$, that cannot create texture?
  (b) Every QST free operation splits into a quantum filter and a residual
  completely positive map (Theorem~\ref{thm:free-normal-form-literature}).
  For any allowed conversion, preparation of a fixed state can replace the
  residual map (Theorem~\ref{thm:exact-convertibility}).
  \corref{cor:conversion-sdp} shows that conversion is possible exactly when an
  allowed filter [\Eref{eq:literature-normal-form}] leaves a positive remainder
  that preparation can supply, and formulates this test as a semidefinite program.
  The blue filter
  and orange preparation contributions add up to the target.}
  \label{fig:qst-conversion-criterion}
\end{figure*} 
\section{Introduction}

Quantum resource theories treat quantum phenomena, such as entanglement, 
as resources for carrying out tasks~\cite{ChitambarGour2019}.
Within this framework,
one can rank states by their usefulness when the available operations
are restricted. If Alice and Bob are on opposite sides of the
world, they can use only local operations and classical communication;
these \emph{free operations} cannot create entanglement between them.
The only way for Alice and Bob to overcome this restriction is by initially
acquiring a \emph{resourceful} state. A shared Bell
pair allows them to perform perfect quantum teleportation, but not every
entangled state does the trick~\cite{BennettTeleportation1993,HorodeckiTeleportation1999}.
For this task, a Bell pair ranks higher than states that do not enable
perfect teleportation.
How can we rank states without reference to a specific task?

A \emph{convertibility} criterion determines if and how states are
ordered independent of any task~\cite{ChitambarGour2019}.
If a free operation converts one state into another, the
source is at least as useful as the target, since it can first be converted
into the target and then used in its place. The two states are equivalent
under free conversion if they can be converted into each other~\cite{GondaSpekkens2023}; in general, a resource
monotone cannot establish this equivalence. A resource monotone
may assign the same value to two states, but the states may not be interconvertible~\cite{Datta2023}.
Hence, a conversion criterion tells us more than individual resource monotones do.

Several resource theories illustrate the importance of exact state
convertibility. For bipartite pure-state entanglement, Nielsen's theorem establishes
a necessary and sufficient majorization criterion for state conversion
under local operations and classical communication
\cite{Nielsen1999}. The same type of ordering appears for pure-state coherence
under incoherent operations \cite{Baumgratz2014,Du2015}. In quantum
thermodynamics, thermomajorization and related families of constraints determine
which nonequilibrium transitions are allowed at the nanoscale, revealing
restrictions that are invisible to the ordinary macroscopic free energy
\cite{HorodeckiOppenheim2013,Gour2015,GourJennings2018}. These examples show how convertibility
can expose the geometry and operational content of a resource theory.

Quantum-state texture (QST) treats the departure of a density matrix from 
a flat state in a fixed basis as a resource~\cite{Parisio2024}.
A zoo of texture measures beyond the original rugosity, along with their 
properties, has been studied to quantify the 
resource~\cite{Wang2025,ZhangWeight2025,Cui2025,Muthuganesan2026,Cao2026,ChenLei2026,Lin2026,deLeonReachability,ChenLei2026,ZhangEstimation2026}.
Notably, connections between QST and many-body physics have been explored 
by investigating quantum criticality~\cite{Patra2026,Celeri2026,Casagrande2026}.
Furthermore, QST has been studied in connection with quantum battery
capacity~\cite{Battery2026}, quantum speed limits~\cite{XieSpeed2026}, and
relativistic settings~\cite{HuangAtoms2025,HuangLorentz2026,ZhangNoninertial2026,HuangSdS2026,ZhangBlackHole2026}.
Returning to theoretical aspects, QST has been extended to set texture~\cite{deLeon2026}
and block texture~\cite{Chu2026}, studied
in the context of fixed-point resource theories~\cite{Greenwood2026} and
resource theories beyond convexity~\cite{Salazar2026}. QST has also been related to other
quantum resources~\cite{Cui2025,Patra2026} and phase-sensitive
superposition~\cite{WangSuperposition2026}.
Remarkably, QST dynamics under quantum channels has been studied 
experimentally for the first time~\cite{Vieira2026}.
Ultimately, QST has attracted interest beyond its original application in the
identification of gates.

State convertibility within QST theory has been partially resolved.
The established criteria include single-qubit conversions and pure-to-mixed
conversions in arbitrary dimension.
First, probabilistic conversion was studied for pure-to-pure and mixed-to-pure
single-qubit transformations~\cite{Cui2025}.
Subsequently, deterministic conversion criteria were established
for arbitrary pairs of single-qubit states and for pure-to-mixed conversions
in arbitrary dimension~\cite{Lin2026}. Single-qubit state conversion 
was also discussed using the general form of single-qubit QST
free operations~\cite{deLeon2026}.
What remains missing is a conversion criterion for mixed states
in arbitrary finite dimension.

Here, we add two missing elements to QST theory that together complete the criterion
for converting one state into another under QST free operations.
First, in Theorem~\ref{thm:free-normal-form-literature}, we derive a normal
form for QST free operations in arbitrary dimensions. Each operation decomposes
into a quantum filter and a residual completely positive map. The filter acts
non-trivially only on the maximally textured subspace, the orthogonal
complement of the unique free state.
The residual map supplies the remaining probability, so both contributions add
up to a quantum state. Second, in Theorem~\ref{thm:exact-convertibility},
we address the state conversion problem, depicted in \Fref{fig:qst-conversion-criterion}(a).
As illustrated in \Fref{fig:qst-conversion-criterion}(b), we prove that 
a state conversion is allowed if and only if there is a QST free operation 
achieving it that admits a form in which the residual map is a
preparation map that supplies the part that the quantum filter misses.
Importantly, we show a number of non-trivial results follow from both theorems,
including the fact that convertibility between two states can be determined by
semidefinite programming.

The article is organized as follows. In \Sref{sec:setup}, we introduce the
conversion problem and review the two scalar constraints that together give
the qubit conversion criterion.
We draw particular attention to Secs.~\ref{sec:texture-weight} and 
\ref{sec:qubit-known}, in which we discuss how the maximal fraction of a state 
in the free state, called here the maximal free fraction, stands on equal
footing with $D$ times the population in the free state, a quantity known as
the grand sum.
In \Sref{sec:free-operations}, we establish the normal form of QST free
operations and use it to recover known grand-sum conservation
results~\cite{Vieira2026}, and extend the Kraus-rank bound for single-qubit QST free
operations~\cite{deLeon2026} to arbitrary dimension.
In \Sref{sec:theorem}, we present the exact state conversion criterion, its
semidefinite formulation, and consequences for universal sources and pure
states. In \Sref{sec:consequences}, we attempt to make the conversion criterion
concrete and present illustrative protocols, explicit
qutrit tests, and the obstruction beyond the two scalar measures that work in 
the single-qubit case.
Finally, in \Sref{sec:conclusions}, we summarize our results and offer an outlook.

\section{Preliminaries}
\label{sec:setup}

We review how state conversion is formulated in QST and what is already
known about it. In \Sref{sec:conversion-problem}, we set up the conversion
problem and write the source and target states in blocks relative to the
textureless state. In \Sref{sec:texture-weight}, we discuss the complementary
constraints imposed by geometric texture and texture weight. Finally, in
\Sref{sec:qubit-known}, we revisit the known qubit criterion and show how these
two constraints together determine which qubit conversions are possible.

\subsection{Conversion problem in quantum-state texture}
\label{sec:conversion-problem}

The theory of QST treats the deviation from the state
\begin{equation}
  \includegraphics[width=.4\columnwidth]{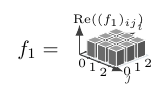}
  \label{eq:free-state}
\end{equation}
as a resource. Mathematically, $f_1=\dyad{f_1}$, with
$\ket{f_1}=\frac{1}{\sqrt D}\sum_{k=0}^{D-1}\ket{k}$, $D$ being the dimension
of the system [$D=3$ in \Eref{eq:free-state}], and $\{\ket{k}\}$ a fixed basis.
This pictorial image of $f_1$
shows that it is textureless, in the sense that its histogram is flat. Any
other state is textured and thus possesses the resource. The textureless state
$f_1$ represents the only free state of this resource theory.

Deterministic state conversion in QST determines whether a free operation
can transform a source state $\rho$ into a target state $\sigma$.
The free operations are completely positive trace-preserving (CPTP)
maps $\Lambda$ satisfying
\begin{equation}
  \Lambda(f_1)=f_1.
  \label{eq:free-map}
\end{equation}
We call these channels QST free operations and write
$\rho\longrightarrow\sigma$ when one of them transforms the input
$\rho$ into the target $\sigma$ with certainty. We seek both a criterion
for the existence of such a channel and a protocol that realizes the
conversion.

Every Kraus operator of a QST free operation maps the textureless
vector to a scalar multiple of itself. Let us write the Kraus representation
of a QST free operation as $\Lambda(X)=\sum_jK_jXK_j^\dagger$.
Then the fixed-state condition $\Lambda(f_1) = f_1$ implies
\begin{equation}
  K_j\ket{f_1}=a_j\ket{f_1}
  \label{eq:fixed-kraus}
\end{equation}
for all Kraus operators $K_j$.
This property serves as the starting point for the normal form of QST free
operations that we derive in \Sref{sec:free-operations}.

The grand sum of a state $X$, $\Sigma(X)$, is defined as the sum of
all matrix elements in the reference basis~\cite{Parisio2024}. For any state $X$,
\begin{equation}
  \Sigma(X):=\sum_{j,k=0}^{D-1}\mel{j}{X}{k}
  =D\mel{f_1}{X}{f_1}=D\Tr(f_1X).
  \label{eq:pq-def}
\end{equation}
Hence $\Sigma(X)/D$ is the probability of finding the system in the
textureless state.

A block decomposition separates the population and coherences associated
with the textureless state from the block on its orthogonal complement, the
maximally textured subspace.
States supported entirely on this subspace are termed maximally textured
states~\cite{deLeonReachability}.
Let us define the complementary projector
\begin{equation}
  f_1^\perp=\id_D-f_1
\end{equation}
and the orthonormal basis $\{\ket{f_1},\ket{e_1},\ldots,\ket{e_{D-1}}\}$,
where $\{\ket{e_a}\}_{a=1}^{D-1}$ spans the maximally textured subspace.
In this basis, the source and target states take the form
\begin{equation}
  \rho=
  \begin{pmatrix}
    \Sigma(\rho)/D & x^\dagger\\
    x & R
  \end{pmatrix},
  \qquad
  \sigma=
  \begin{pmatrix}
    \Sigma(\sigma)/D & y^\dagger\\
    y & S
  \end{pmatrix}.
  \label{eq:block-states}
\end{equation}
Here $x,y\in\C^{D-1}$ contain the coherences with the textureless subspace,
and $R,S\succeq0$ are the blocks on the maximally textured subspace, with
$\Tr R=1-\Sigma(\rho)/D$ and $\Tr S=1-\Sigma(\sigma)/D$.
The notation $\succeq0$ denotes positive semidefiniteness. These blocks
will describe both the action of a free channel and the constraints on
a conversion.

\subsection{Geometric texture and texture weight}
\label{sec:texture-weight}

Texture measures quantify the resource in QST theory.
A texture measure $\mathcal T$ is defined
as a function that vanishes on $f_1$ and does not increase under QST
free operations. Immediately, it
follows that
\begin{equation}
  \rho\longrightarrow\sigma
  \quad\Longrightarrow\quad
  \mathcal T(\sigma)\le\mathcal T(\rho).
  \label{eq:conversion-monotone-constraint}
\end{equation}
That is, conversion implies the source must be assigned at least the same
texture value as the target, for any texture measure.
Whether these necessary conditions are sufficient depends on the
states and the dimension.

Two known texture measures provide scalar constraints on conversion.
Geometric texture is the probability of not finding the
system in the
textureless state~\cite{Wang2025}:
\begin{equation}
  \mathcal T_g(\rho)=1-\mel{f_1}{\rho}{f_1}
  =1-\frac{\Sigma(\rho)}{D}.
  \label{eq:geometric-texture}
\end{equation}
Its monotonicity under QST free operations constrains
the grand sum to be nondecreasing.
The same conversion constraint follows from any texture monotone that is
a strictly decreasing function of the grand sum, e.g., the rugosity~\cite{Parisio2024}.
Texture weight measures the smallest residual fraction left after
separating a free component from the state
\cite{ZhangWeight2025,ChenLei2026,Lin2026}:
\begin{equation}
  \mathcal T_w(\rho)
  =\min_{\substack{0\le s\le1\\ \tau\in\mathcal D(\mathcal H)}}
    \{s:\rho=(1-s)f_1+s\tau\},
  \label{eq:texture-weight}
\end{equation}
where $\mathcal D(\mathcal H)$ denotes the set of density operators on the
$D$-dimensional Hilbert space $\mathcal H$ of the system.
Although a plethora of texture measures have been introduced,
the geometric texture and texture weight measures truly capture
complementary aspects of state texture,
as we discuss in the remainder of this section
and the next section.

Turning the definition of texture weight in \Eref{eq:texture-weight} around
amounts to asking for the
\emph{maximal free-state fraction} of a state.
This then allows us to
find a complementary relation between the grand sum and the maximal free fraction.
Let us define the maximal free-state fraction of $\rho$:
\begin{equation}
  c_\rho=\max\{t\in[0,1]:\rho-tf_1\succeq0\},
  \label{eq:max-free-fraction}
\end{equation}
such that
\begin{equation}
  \mathcal T_w(\rho)=1-c_\rho.
  \label{eq:weight-schur}
\end{equation}
The maximal free-state fraction can be calculated
by invoking the generalized Schur-complement criterion, which establishes that
\begin{align}
  \rho-tf_1\succeq0
  &\quad\Longleftrightarrow\quad t\le c_\rho.
  \label{eq:free-fraction-positivity}
\end{align}
In turn, this implies that
\begin{align}
  c_\rho&=\frac{\Sigma(\rho)}{D}-x^\dagger R^+x,
  \label{eq:schur-free-fraction}
\end{align}
where $R^+$ is the Moore--Penrose inverse, equal to $R^{-1}$
when $R$ is positive definite.
Indeed, \Eref{eq:schur-free-fraction} tells us that the maximal free
fraction is the generalized Schur complement
of the block $R$ of the state $\rho$.
The same Schur-complement quantity appears in low-temperature
Gibbs-preserving thermodynamics, where the distinguished pure state is
the ground state \cite{NarasimhacharGour2015}.

We stress that the population $\Sigma(\rho)/D$ and the fraction $c_\rho$ describe
different aspects of the free state: the former is a measurement
probability, while the latter is a weight in a convex decomposition.
They obey $0\le c_\rho\le\Sigma(\rho)/D\le1$. Monotonicity of texture
weight requires $c_\sigma\ge c_\rho$ for an allowed conversion, whereas
geometric texture requires $\Sigma(\sigma)\ge\Sigma(\rho)$. In
\Sref{sec:theorem}, we recover both conditions directly from the
conversion protocol.

Texture weight distinguishes the textureless state from every textured
pure state, but does not distinguish among textured pure states.
Indeed, $c_{f_1}=1$ and $\mathcal T_w(f_1)=0$, whereas every textured
pure state has $c_\rho=0$ and $\mathcal T_w(\rho)=1$.

\subsection{Known qubit-state conversion criteria}
\label{sec:qubit-known}

Deterministic criteria are known for pure sources in any finite dimension
and arbitrary sources for qubits. A pure state $\rho$ can reach any
target $\sigma$ if and only if $\Sigma(\sigma)\ge\Sigma(\rho)$~\cite{Lin2026}.
For arbitrary mixed single-qubit sources, a complete criterion is
known~\cite{Lin2026,deLeon2026}. We use this section to revisit the criterion
of Ref.~\cite{Lin2026} in light of the reformulation of the texture weight
discussed above.

\begin{figure}[t]
  \centering
  \includegraphics[width=\columnwidth]{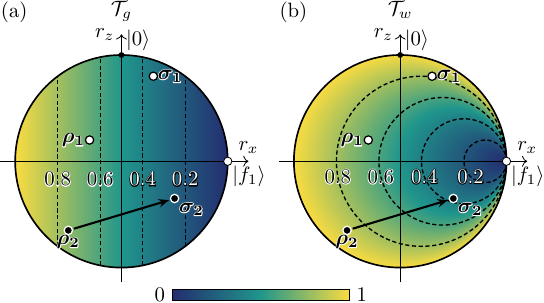}
  \caption{(a) Geometric texture [\Eref{eq:geometric-texture}] and
  (b) texture weight [Eqs.~\eqref{eq:texture-weight} and \eqref{eq:weight-schur}]
  for single-qubit states in the
  $xz$ plane of the Bloch sphere. The equitexture curves are shown
  as dashed lines
  at $0.2$, $0.4$, $0.6$, and $0.8$. The state conversion criterion in
  \Eref{eq:qubit-two-measures} rules out conversion from $\rho_1$ to
  $\sigma_1$ and allows conversion from $\rho_2$ to $\sigma_2$.}
  \label{fig:bloch-textures}
\end{figure}

Single-qubit convertibility is completely determined by geometric texture
and texture weight.
Rewriting the conditions of Ref.~\cite{Lin2026}, originally stated using
the Bloch components of the source and target states, in terms of grand
sums and determinants yields~\cite{Lin2026}:
\begin{equation}
  \rho\longrightarrow\sigma
  \quad\Longleftrightarrow\quad
  \begin{aligned}
    \Sigma(\sigma)&\ge\Sigma(\rho),\\
    \qty[2-\Sigma(\rho)]\det\sigma
    &\ge
    \qty[2-\Sigma(\sigma)]\det\rho.
  \end{aligned}
  \label{eq:qubit-det-form}
\end{equation}
The first inequality is the geometric-texture condition.
The second, we show is precisely the texture-weight condition.
For a textured single-qubit state, the block on the maximally textured
subspace is the positive scalar
$1-\Sigma(\rho)/2$, since $\Tr(\rho) = \Tr(R) + \Sigma(\rho)/2 = 1$,
and \Eref{eq:schur-free-fraction} becomes
\begin{equation}
  c_\rho=\frac{\det\rho}{1-\Sigma(\rho)/2}
  =\frac{\Sigma(\rho)}{2}
   -\frac{\abs{x}^2}{1-\Sigma(\rho)/2}.
  \label{eq:qubit-c}
\end{equation}
Using this expression in \Eref{eq:qubit-det-form}, we obtain
\begin{equation}
  \rho\longrightarrow\sigma
  \quad\Longleftrightarrow\quad
  \begin{aligned}
    \mathcal T_g(\sigma)&\le\mathcal T_g(\rho),\\
    \mathcal T_w(\sigma)&\le\mathcal T_w(\rho).
  \end{aligned}
  \label{eq:qubit-two-measures}
\end{equation}
We thus recover the QST version of the two-monotone qubit criterion for
channels that fix a pure state~\cite{NarasimhacharGour2015}.
We can see both constraints at work in \Fref{fig:bloch-textures}.
The states $\rho_1$ and $\sigma_1$ are incomparable, while the conversion
$\rho_2\longrightarrow\sigma_2$ is allowed.

The two measures constrain qubit conversion in different ways. For a
textured pure source, $\mathcal T_w(\rho)=1$, so only the grand-sum
condition restricts the target. For a mixed qubit source, the maximal
free fraction satisfies $c_\rho>0$. Since this fraction cannot decrease
under a free operation, such a source cannot reach a textured pure state,
whose maximal free fraction is zero. Ref.~\cite{deLeon2026} gives examples
of incomparable mixed qubit states.

What happens when the source is mixed and the dimension is greater than two?
The blocks $R$ and $S$ on the maximally textured subspace
[\Eref{eq:block-states}] are then matrices, and the two scalar measures
need not capture all restrictions on their populations and coherences.
In the next section, we characterize the QST free operations. We then use
their structure to obtain an exact conversion criterion.

\section{QST free operations}
\label{sec:free-operations}

QST free operations possess a known block-triangular Kraus structure due to the
fact that they leave the pure textureless state unchanged. This
block-triangular form follows from general results on invariant subspaces for
quantum channels~\cite{CirilloTicozzi2015}. In the case of low-temperature
Gibbs-preserving operations, Ref.~\cite{NarasimhacharGour2015} wrote down the
Kraus operators in this form along with the conditions of trace preservation.
When the Gibbs state is a rank-one projector, those operations coincide with
QST free operations up to a unitary change of basis. The same structure also
appears in the study of generalized texture and of pure fixed-point resource
theories~\cite{Greenwood2026}.

From this Kraus structure, we derive a normal form for QST free operations in
arbitrary finite dimensions. Every QST free operation decomposes into a quantum
filter and a residual completely positive map. Here, a quantum filter
corresponds to one measurement outcome, with unnormalized output
$A\rho A^\dagger$ and $A^\dagger A\preceq\id_D$~\cite{Verstraete2001Filtering}.
It is important to note that only the filter depends on the coherences of the
input state with the textureless state, whereas the residual map acts only on
the block of the input state in the maximally textured subspace. In
\Sref{sec:theorem}, we use this decomposition to reduce state conversion to the
problem of finding a filter whose output can be completed to the target state.
In the remainder of this section, we first present the normal form and prove
its uniqueness; we then use it in order to characterize grand-sum preservation
and to find upper bounds on the Kraus rank. We denote by $M_n(\C)$
the space of complex $n\times n$ matrices and by $M_{m,n}(\C)$ the space of
complex $m\times n$ matrices.

\begin{theorem}[Normal form of QST free operations]
\label{thm:free-normal-form-literature}
A CPTP map $\Lambda:M_D(\C)\rightarrow M_D(\C)$
is a QST free operation if and only if its action on every $X\in M_D(\C)$ 
can be written as
\begin{subequations}
\label{eq:literature-normal-form}
\begin{equation}
  \Lambda(X)=AXA^\dagger+\Phi(f_1^{\perp}Xf_1^{\perp}),
  \label{eq:literature-channel}
\end{equation}
where $A=\operatorname{diag}(1,T)$ in the block decomposition of
\Eref{eq:block-states}. The matrix $T\in M_{D-1}(\C)$ and the completely
positive map $\Phi:M_{D-1}(\C)\rightarrow M_D(\C)$ satisfy
\begin{align}
  T^\dagger T&\preceq\id_{D-1},
  \label{eq:literature-contraction}\\
  \Phi^\dagger(\id_D)&=\id_{D-1}-T^\dagger T.
  \label{eq:literature-defect}
\end{align}
\end{subequations}
In the argument of $\Phi$, we identify $f_1^{\perp}Xf_1^{\perp}$ with
the $(D-1)\times(D-1)$ lower-right block of $X$.
We use $\Phi^\dagger$ for the Hilbert--Schmidt adjoint, defined by
$\Tr[Y\Phi(Z)]=\Tr[\Phi^\dagger(Y)Z]$.
\end{theorem}

We prove Theorem~\ref{thm:free-normal-form-literature} in
Appendix~\ref{app:normal-form}. Let us now examine how each of the two terms
acts on the blocks of a state.

The filter preserves the textureless state and can rotate and attenuate
components in the maximally textured subspace. If we write the state as in
\Eref{eq:block-states}, we obtain
\begin{equation}
  \Lambda(\rho)=
  \begin{pmatrix}
    \Sigma(\rho)/D&x^\dagger T^\dagger\\
    Tx&TRT^\dagger
  \end{pmatrix}+\Phi(R).
  \label{eq:literature-block-action}
\end{equation}
The first term is the unnormalized output of the filter. We see that the
coherences with the textureless subspace transform as $x\mapsto Tx$ and
the block on the maximally textured subspace as $R\mapsto TRT^\dagger$.
By the contraction condition in \Eref{eq:literature-contraction}, the filter
cannot increase either the norm of $x$ or the total population in the maximally
textured subspace. However, particular coherences inside the maximally textured
subspace can still increase, due to the fact that the filter can mix different
directions of that subspace.

On the other hand, the residual term $\Phi(R)$ supplies the remaining
probability so that the output state is normalized. Its argument is only the
block $R$, although $\Phi(R)$ can contribute to all blocks of the output
density matrix, including the coherences. Using \Eref{eq:literature-defect},
one obtains
$\Tr[\Phi(R)]=\Tr[(\id_{D-1}-T^\dagger T)R]=1-\Tr(A\rho A^\dagger)$.
Therefore, $T$ fixes the filter as well as the trace that the residual map has
to supply, whereas $\Phi$ determines how that trace is distributed among the
blocks of the output state.

We stress that, once the adapted basis is fixed, the pair $(T,\Phi)$ is
uniquely determined by the channel $\Lambda$, independent of the chosen Kraus
representation. To verify this, let us consider two pairs $(T_1,\Phi_1)$ and
$(T_2,\Phi_2)$ that represent the same channel $\Lambda$. For any vector
$\ket{v}$ in the maximally textured subspace, we have
$\ket{T_1v}=\Lambda(\ket{v}\bra{f_1})\ket{f_1}=\ket{T_2v}$.
Because $T_1$ and $T_2$ coincide on every vector of their domain, it follows
immediately that $T_1=T_2$, which in turn fixes $A=\operatorname{diag}(1,T)$.
Then, in order to recover the residual map, we subtract the contribution of
the filter. For every operator $Z=f_1^{\perp}Zf_1^{\perp}$, we obtain
$\Phi_1(Z)=\Lambda(Z)-AZA^\dagger=\Phi_2(Z)$, and hence $\Phi_1=\Phi_2$.

In the case of a single qubit, every QST free operation consists of a diagonal
filter and a residual branch that prepares a fixed state from the population
in $\ket{-}$. In the adapted basis $(\ket{+},\ket{-})$,
\Eref{eq:literature-channel} takes the form
$\Lambda(\rho)=A\rho A^\dagger+\rho_{22}B$. By comparing this with the Choi
matrix in Eq.~(21) of de Leon \textit{et al.}~\cite{deLeon2026}, one obtains
$A=\operatorname{diag}(1,\bar\lambda)$ and
\[
  B=\Phi(1)=\begin{pmatrix}
    1-\eta & q\\
    \bar q & \eta-\abs{\lambda}^2
  \end{pmatrix}.
\]
The positivity condition $B\succeq0$ is equivalent to the bounds in their
Eq.~(23), and $\Tr(B)=1-\abs{\lambda}^2$. When $\Tr(B)>0$, the residual branch
prepares the state $B/\Tr(B)$ with probability $\rho_{22}\Tr(B)$. In Bloch
coordinates $(x,y,z)$, writing $w=z+iy$, the channel gives
$x'=1-\eta+\eta x$ and $w'=\lambda w+q(1-x)$, recovering their Eq.~(28). The
filter rotates and contracts each disk at fixed $x$, while the residual branch
shifts the center of the disk by $q(1-x)$ in the transverse plane.

We now consider two simple examples of the normal form, one in which there is
no contribution from $\Phi$, and another that transfers population to the
textureless state.

\paragraph*{Unitary QST free operations.---}
Let us take $T=V$ to be a unitary operator on the maximally textured subspace.
Then $\id_{D-1}-T^\dagger T=0$, which implies $\Phi=0$, and the channel reduces
to $\Lambda(X)=U_f XU_f^\dagger$, with $U_f=\operatorname{diag}(1,V)$. These
are all the unitary channels that leave the textureless state unchanged.

\paragraph*{Partial reset.---}
A partial reset transfers a fraction of the population in the maximally
textured subspace to the textureless state, controlled by a parameter
$0\le\eta\le1$. Let us choose $T=\sqrt{\eta}\,\id_{D-1}$ and
$\Phi(Z)=(1-\eta)\Tr(Z)f_1$. For a state written in the block form of
\Eref{eq:block-states}, we obtain
\begin{equation}
  \Lambda_\eta(\rho)=
  \begin{pmatrix}
    1-\eta+\eta\Sigma(\rho)/D & \sqrt{\eta}\,x^\dagger\\
    \sqrt{\eta}\,x & \eta R
  \end{pmatrix}.
\end{equation}
Here, $\eta$ represents the fraction of the initial population that remains in
the maximally textured subspace, while the coherences with the textureless
state are attenuated by $\sqrt{\eta}$. 
The rest of the population then moves from that subspace to the textureless state. 
At $\eta=1$ the partial reset QST free operation is the identity; at $\eta=0$ it is the
complete reset $\Lambda_0(X)=\Tr(X)f_1$.

For a general free channel, only the residual map transfers population
into the textureless subspace. Since
$\Sigma(\rho)=D\Tr(f_1\rho)$, conservation of this population is
equivalent to grand-sum preservation. By taking the adjoint
of \Eref{eq:literature-channel} and using
$A^\dagger f_1A=f_1$, we find
\begin{equation}
  \Lambda^\dagger(f_1)=f_1\oplus\Phi^\dagger(f_1).
  \label{eq:free-dual-reference}
\end{equation}
If we take the expectation in the input state and multiply by $D$, we obtain
\begin{equation}
  \Sigma[\Lambda(\rho)]
  =\Sigma(\rho)+D\Tr[\Phi^\dagger(f_1)R]\ge\Sigma(\rho),
  \label{eq:grand-sum-increment}
\end{equation}
where the inequality follows from positivity. Thus, all population
transferred into the textureless subspace comes from the action of the
residual map on $R$, whereas the input coherences $x$ do not contribute to the
increase of the grand sum.

Preserving the grand sum for every input state is equivalent to requiring
$\Lambda^\dagger(f_1)=f_1$, as noted in Ref.~\cite{Vieira2026}.
This follows directly from the adjoint identity
$\Sigma[\Lambda(\rho)]=D\Tr[\Lambda^\dagger(f_1)\rho]$.
In terms of our normal form, this condition means that the output of the
residual map must be supported entirely on the maximally textured subspace.

\begin{corollary}[Grand-sum-preserving QST free operations]
\label{cor:grand-sum-preservation}
A QST free operation is grand-sum-preserving if and only if
$\Phi^\dagger(f_1)=0$. Equivalently, it admits
a Kraus representation
\begin{equation}
  K_0=\begin{pmatrix}1&0\\0&T\end{pmatrix},
  \qquad
  K_\mu=\begin{pmatrix}0&0\\0&B_\mu\end{pmatrix},
  \label{eq:preserving-kraus-form}
\end{equation}
where $B_\mu$ acts on the maximally textured subspace and
\begin{equation}
  T^\dagger T+\sum_\mu B_\mu^\dagger B_\mu=\id_{D-1}.
  \label{eq:preserving-kraus-normalization}
\end{equation}
Its action on a state is therefore
\begin{equation}
  \Lambda(\rho)=
  \begin{pmatrix}
    \Sigma(\rho)/D & x^\dagger T^\dagger\\
    Tx & TRT^\dagger+\sum_\mu B_\mu RB_\mu^\dagger
  \end{pmatrix}.
  \label{eq:preserving-block-action}
\end{equation}
\end{corollary}
\begin{proof}
From \Eref{eq:grand-sum-increment}, we see that preserving the grand sum
for every input state is equivalent to $\Phi^\dagger(f_1)=0$, since $R$ can
represent any maximally textured state.

To obtain the equivalent Kraus form, we restrict the outputs of the residual
map to the maximally textured subspace. Let us write
$\Phi(Z)=\sum_\mu C_\mu ZC_\mu^\dagger$, where each $C_\mu$ maps the
maximally textured subspace into the system's full Hilbert space. Then
$\Phi^\dagger(f_1)=\sum_\mu C_\mu^\dagger f_1C_\mu$. Since each term in this
sum is positive semidefinite, the sum vanishes if and only if $f_1C_\mu=0$ for
every $\mu$. The output of every $C_\mu$ is then supported in the maximally
textured subspace. If we denote their nonzero blocks by $B_\mu$, we obtain
\Eref{eq:preserving-kraus-form}, and the trace-preservation condition yields
\Eref{eq:preserving-kraus-normalization}.
Conversely, any channel with Kraus operators of this form preserves
$\Sigma(\rho)$ for every input state, as one can see directly from
\Eref{eq:preserving-block-action}.
\end{proof}

The block-diagonal form of the Kraus operators does not allow population 
to move between the textureless and maximally textured subspaces.
Moreover, the Kraus operators in \Eref{eq:preserving-kraus-form} commute with
$f_1$, since $f_1=\operatorname{diag}(1,0,\ldots,0)$ in the same basis.
This property is preserved under any change of Kraus representation,
because each new Kraus operator is a linear combination of the original ones.

We want to stress our reasons for choosing \emph{grand-sum-preserving} over
QST-preserving, the term used for these free operations in Ref.~\cite{Vieira2026}.
Preserving the populations of the textureless and maximally textured
subspaces fixes the geometric texture,
$\mathcal T_g(\rho) = 1 - \Sigma(\rho)/D$.
In fact, the grand sum fixes the state's symmetric representative on the
\emph{isotropic axis}~\cite{deLeonReachability}, and every texture monotone
that depends only on $\Sigma(\rho)$ is therefore preserved.
Other texture monotones can still decrease. Let us consider the dephasing channel
$X\mapsto f_1Xf_1+f_1^{\perp}Xf_1^{\perp}$. It preserves the grand sum
but sends any pure state with $0<\Sigma(\rho)<D$ to a block-diagonal
state $\sigma$ whose texture weight is $\mathcal T_w(\sigma)=1-\Sigma(\rho)/D$
[see \Eref{eq:weight-schur}],
down from $1$. The drop in texture weight rules out a conversion under
QST free operations back to the original pure state. Therefore, we would reserve
the term QST-preserving for a free operation whose action on a given textured 
state can be reversed by another free operation. Two examples of such channels
are free unitaries and the channels in
\Eref{eq:replacement-free-channel} when used to convert between maximally
textured states.

Unitality, $\Lambda(\id_D)=\id_D$, implies grand-sum preservation
in every dimension, while the converse holds only for qubits.

\begin{corollary}[Unital QST free operations are grand-sum preserving.]
\label{cor:unital-free}
If a QST free operation is unital, then it preserves the grand sum.
The converse also holds for a single qubit.
\end{corollary}
\begin{proof}
From \Eref{eq:literature-channel}, unitality imposes
$\Phi(\id_{D-1})=\id_D-AA^\dagger$, whose support lies in
the maximally textured subspace. Hence,
$\Tr[\Phi^\dagger(f_1)]=\Tr[f_1\Phi(\id_{D-1})]=0$.
Since $\Phi^\dagger(f_1)\succeq0$, it follows that $\Phi^\dagger(f_1)=0$, and
\corref{cor:grand-sum-preservation} implies that the grand sum is preserved.

In the single-qubit case, grand-sum preservation also implies unitality because
every Kraus operator in the representation of
Corollary~\ref{cor:grand-sum-preservation} is diagonal. Since $T=t$ and
$B_\mu=b_\mu$ are scalars, \Eref{eq:preserving-kraus-normalization} reduces to
$\abs{t}^2+\sum_\mu\abs{b_\mu}^2=1$, which also implies
$K_0K_0^\dagger+\sum_\mu K_\mu K_\mu^\dagger=\id_2$,
the Kraus condition for unitality.
\end{proof}
For $D\ge3$, nonunital dynamics without population transfer to 
the textureless subspace can occur within the maximally textured subspace.
We can see this with a measure-and-prepare channel.
\paragraph*{Measurement and preparation.---}
A measurement that distinguishes the textureless subspace from the maximally
textured subspace, followed by state preparation, defines a QST free operation.
In \Eref{eq:literature-normal-form}, let us take the filter $A=f_1$ and the
residual map $\Phi(Z)=\Tr(Z)\omega$ for an arbitrary state $\omega$.
This yields
\begin{equation}
  \Lambda_\omega(X)=\Tr(f_1X)f_1+\Tr(f_1^{\perp}X)\omega.
  \label{eq:replacement-free-channel}
\end{equation}
This channel measures whether the input system is found in the textureless
or maximally textured subspace and prepares $f_1$ or $\omega$, respectively.
When $\omega$ is restricted to a maximally textured state, this map was used in
Ref.~\cite{MaFourier2026} to explain why some states retain maximal
texture under Hawking radiation. Ref.~\cite{deLeonReachability} uses the general
form to show that all maximally textured states must attain the same global
maximum for any texture monotone.

Notice that for $D\ge3$ this measure-and-prepare channel can preserve the grand
sum without being unital. Indeed, if we set $\omega=\ket{e_1}\bra{e_1}$ with
$\ket{e_1}$ in the maximally textured subspace, then no population crosses
between the two subspaces, so the grand sum does not change. Even so,
$\Lambda_\omega(\id_D)=f_1+(D-1)\omega\ne\id_D$ whenever $D\ge3$.
So grand-sum preservation does not rule out nonunital dynamics happening purely
inside the maximally textured subspace.

Once again, we emphasize that the result of Corollary~\ref{cor:unital-free},
as well as that of Corollary~\ref{cor:grand-sum-preservation}, concerns
grand-sum preservation under QST free operations.
For that reason, we prefer calling such operations \emph{grand-sum-preserving} 
because \emph{QST-preserving} can misleadingly suggest that every texture 
monotone is preserved.
The dephasing channel discussed before Corollary~\ref{cor:unital-free} is a unital
QST free operation that preserves the grand sum, hence the geometric texture,
but can decrease the texture weight. This is simply a consequence of the
distinct and complementary roles of the grand sum and the maximal free fraction
[see \Eref{eq:max-free-fraction}].

The same restrictions on the residual map also give us upper bounds on the
Kraus rank (the smallest number of operators needed in a Kraus representation
of the channel). In general, the Kraus operators of $\Phi$ map the maximally
textured subspace into the system's full Hilbert space; when the channel
preserves the grand sum, their outputs stay inside the maximally textured
subspace. Simply counting the dimensions of these two operator spaces extends
the known single-qubit maximum of three~\cite{deLeon2026} to any dimension $D$,
and gives a smaller maximum when the grand sum is conserved.

\begin{corollary}[Kraus-rank bounds]
\label{cor:kraus-ranks}
Every QST free operation admits a Kraus representation with at most 
$1+D(D-1)$ Kraus operators. If it is grand-sum preserving, then the free operation
has Kraus rank at most $1+(D-1)^2$.
Both bounds are attained by channels in
\Eref{eq:replacement-free-channel}.
\end{corollary}
\begin{proof}
Both bounds come directly from counting linearly independent Kraus operators.
Apart from $A$, the residual map $\Phi$ is built from operators $C_\mu$ mapping
the maximally textured subspace into the full Hilbert space. Since the space of
such operators has dimension $D(D-1)$, we never need more than $D(D-1)$
linearly independent operators to represent $\Phi$. When the grand sum is
preserved, Corollary~\ref{cor:grand-sum-preservation} forces the outputs of the
$C_\mu$ to lie inside the maximally textured subspace, which drops the
dimension to $(D-1)^2$. Adding the single operator $A$ gives the two upper
bounds.

To show that measurement-and-preparation channels actually saturate both
bounds, let us take a state $\omega=\sum_{k=1}^m w_k\ket{u_k}\bra{u_k}$ of
rank $m$, where $w_k>0$ and the vectors $\ket{u_k}$ are orthonormal. Picking an
orthonormal basis $\{\ket{e_a}\}_{a=1}^{D-1}$ of the maximally textured
subspace, the channel in \Eref{eq:replacement-free-channel} can be written with
Kraus operators $f_1$ and $\sqrt{w_k}\ket{u_k}\bra{e_a}$. All $1+m(D-1)$ operators are linearly independent, so no representation with
fewer Kraus operators exists. Taking $\omega=\id_D/D$ yields $m=D$, and taking
$\omega=f_1^{\perp}/(D-1)$ yields $m=D-1$; both bounds are therefore attained.
\end{proof}
For a single qubit ($D=2$), these maximum Kraus ranks are $3$ and $2$, whereas
for a qutrit ($D=3$) they are $7$ and $5$. In the case of a single qubit, the
bound of $3$ recovers the result of Ref.~\cite{deLeon2026}, while the bound of
$2$ comes from the additional condition of grand-sum preservation.

In the next section, we use the normal form of QST free operations in order to
study deterministic state conversion in arbitrary finite dimensions.

\section{Exact convertibility in arbitrary dimension}
\label{sec:theorem}

By means of the normal form of \Sref{sec:free-operations}, we now address the
state conversion problem for arbitrary finite dimensions. For a given pair of
states $\rho$ and $\sigma$, the residual map $\Phi$ contributes only one
positive semidefinite operator $\Phi(R)$ to the output. First, we prove that
this contribution can always be replaced by the preparation of a fixed state
$\omega$, while keeping the same quantum filter. In turn, this reduction gives
a state conversion criterion formulated as a semidefinite program, from which
we recover necessary scalar conditions and obtain explicit conversion protocols
for several classes of states.

\begin{theorem}[Exact deterministic convertibility]
\label{thm:exact-convertibility}
The transformation $\rho\longrightarrow\sigma$ is possible under
QST free operations if and only if it can be achieved by a QST free
operation whose residual map in the normal form of
Theorem~\ref{thm:free-normal-form-literature} satisfies
\begin{equation}
  \Phi(Z)=\Tr[(\id_{D-1}-T^\dagger T)Z]\,\omega
  \label{eq:conversion-residual}
\end{equation}
for every $Z\in M_{D-1}(\C)$, where $\omega$ is a density operator
that may depend on the chosen conversion but is independent of $Z$.
\end{theorem}

\begin{proof} 
Let us assume that a QST free operation $\Lambda$ transforms $\rho$ into
$\sigma$. By Theorem~\ref{thm:free-normal-form-literature}, we have
$\sigma=A\rho A^\dagger+\Psi(R)$, where $R=f_1^\perp\rho f_1^\perp$ and $\Psi$
is the residual map of $\Lambda$. From trace preservation, we obtain
$p_{\mathrm{prep}}:=\Tr[\Psi(R)]=\Tr[(\id_{D-1}-T^\dagger T)R]$.

If $p_{\mathrm{prep}}>0$, we define the state $\omega=\Psi(R)/p_{\mathrm{prep}}$.
If instead $p_{\mathrm{prep}}=0$, positivity implies $\Psi(R)=0$, and we can
take $\omega$ to be any density operator.

In either case, the map $\Phi(Z)=\Tr[(\id_{D-1}-T^\dagger T)Z]\,\omega$
satisfies $\Phi(R)=\Psi(R)$. Since $T$ is a contraction,
$\id_{D-1}-T^\dagger T\succeq0$; hence, $\Phi$ is completely positive and
fulfills the normalization condition of
Theorem~\ref{thm:free-normal-form-literature}. Therefore, the channel
$\widetilde\Lambda(X)=AXA^\dagger+\Phi(f_1^\perp X f_1^\perp)$ is a QST free
operation that transforms $\rho$ into $\sigma$.

Conversely, if there exists a QST free operation of the form above that maps
$\rho$ to $\sigma$, then immediately $\rho\longrightarrow\sigma$.
\end{proof} 

From Theorem~\ref{thm:exact-convertibility}, it follows that every allowed
state conversion can be realized by a measurement with two outcomes followed by
conditional preparation. We consider the measurement operators
$A=\operatorname{diag}(1,T)$ and $B=\sqrt{\id_D-A^\dagger A}$.
After the first outcome, we retain the filtered state produced by $A$; after
the second outcome, we discard the system and prepare the fixed state $\omega$.
Taking the average over both outcomes gives the channel
\begin{equation}
  \Lambda(X)=AXA^\dagger+\Tr[(\id_D-A^\dagger A)X]\,\omega.
  \label{eq:filter-reprepare-channel}
\end{equation}
Such a protocol is deterministic because both outcomes are kept, and its
averaged output satisfies $\Lambda(\rho)=\sigma$. Furthermore, it leaves the
textureless state $f_1$ unchanged even when $\omega$ is a textured state. This
comes from the fact that $A\ket{f_1}=\ket{f_1}$, which implies that the second
outcome occurs with probability zero on $f_1$, so $\omega$ is never prepared
for that input.

Therefore, state conversion reduces to finding a quantum filter whose
unnormalized output leaves a positive semidefinite remainder when subtracted
from the target state, that is, $\Delta=\sigma-A\rho A^\dagger\succeq0$. In
the adapted basis of \Eref{eq:block-states}, this remainder has the block form
\begin{equation}
  \Delta=
  \begin{pmatrix}
    \frac{\Sigma(\sigma)-\Sigma(\rho)}{D} & (y-Tx)^\dagger\\
    y-Tx & S-TRT^\dagger
  \end{pmatrix}.
  \label{eq:Delta}
\end{equation}
We stress that positivity of $\Delta$ is the only extra condition needed.
Indeed, $\Tr\Delta$ equals the probability of the preparation branch, and
normalizing $\Delta$ determines $\omega$ whenever $\Tr\Delta>0$. Eliminating
$\omega$ in this way leads to the following feasibility criterion in terms of
$T$.

\begin{corollary}[Semidefinite conversion criterion]
\label{cor:conversion-sdp}
The conversion $\rho\longrightarrow\sigma$ is possible if and only if
there exists $T\in M_{D-1}(\C)$ such that
\begin{equation}
  T^\dagger T\preceq\id_{D-1},
  \qquad
  \Delta\succeq0.
  \label{eq:main-theorem}
\end{equation}
Equivalently, with $A=\operatorname{diag}(1,T)$, the following
semidefinite program (SDP) is feasible:
\begin{subequations}
\label{eq:conversion-sdp}
\begin{align}
  \begin{pmatrix}
    \id_{D-1} & T^\dagger\\
    T & \id_{D-1}
  \end{pmatrix}&\succeq0,
  \label{eq:sdp-contraction}\\
  \begin{pmatrix}
    \sigma & A\rho^{1/2}\\
    \rho^{1/2}A^\dagger & \id_D
  \end{pmatrix}&\succeq0.
  \label{eq:sdp-remainder}
\end{align}
\end{subequations}
\end{corollary}
\begin{proof}
If $\rho\longrightarrow\sigma$, Theorem~\ref{thm:exact-convertibility} yields a
channel of the form in \Eref{eq:filter-reprepare-channel}, for which
$\Delta=p_{\mathrm{prep}}\omega\succeq0$, with
$p_{\mathrm{prep}}=\Tr[(\id_D-A^\dagger A)\rho]$.
Conversely, let us assume that \Eref{eq:main-theorem} holds. Then
$\Tr\Delta=1-\Tr(A\rho A^\dagger)=p_{\mathrm{prep}}$. For
$p_{\mathrm{prep}}>0$, we take $\omega=\Delta/p_{\mathrm{prep}}$; for
$p_{\mathrm{prep}}=0$, positivity forces $\Delta=0$ and any density operator
$\omega$ can be chosen. Substituting this state into
\Eref{eq:filter-reprepare-channel} produces a QST free operation achieving
$\rho\longrightarrow\sigma$.

Taking the Schur complements of the identity blocks in
Eqs.~\eqref{eq:sdp-contraction} and \eqref{eq:sdp-remainder}, one obtains
$\id_{D-1}-T^\dagger T\succeq0$ and $\sigma-A\rho A^\dagger\succeq0$,
respectively. Since both block matrices are affine in the real and imaginary
parts of $T$, these matricial constraints define an SDP feasibility problem.
\end{proof}

On the one hand, this SDP decides whether the conversion
$\rho\longrightarrow\sigma$ is allowed; on the other, any feasible matrix $T$
gives the measurement operator $A$ and the remainder $\Delta$ needed to
construct the conversion protocol. It must be noted that
\Eref{eq:conversion-sdp} does not involve any matrix inverse, so the criterion
also applies when $\rho$ or $\sigma$ has one or more zero eigenvalues.
Moreover, before solving the SDP, two scalar conditions can already rule out a
conversion, because the protocol cannot decrease either the population in the
textureless state or the maximal free fraction.

\begin{corollary}[Necessary conversion conditions]
\label{cor:necessary-conversion}
Every conversion $\rho\longrightarrow\sigma$ under QST free operations
satisfies
\begin{align}
  \Sigma(\sigma)&\ge\Sigma(\rho),
  \label{eq:p-monotone}\\
  c_\sigma&\ge c_\rho.
  \label{eq:free-fraction-monotone}
\end{align}
\end{corollary}
\begin{proof}
Because $A^\dagger\ket{f_1}=\ket{f_1}$, the filter leaves the population in the
textureless state invariant, and hence
\begin{equation}
  \frac{\Sigma(\sigma)-\Sigma(\rho)}{D}
  =p_{\mathrm{prep}}\mel{f_1}{\omega}{f_1}\ge0.
\end{equation}
For the second inequality, the definition of $c_\rho$ in
\Eref{eq:max-free-fraction} gives $\rho-c_\rho f_1\succeq0$. Since
$Af_1A^\dagger=f_1$ and $\Delta\succeq0$, we obtain
\begin{equation}
  \sigma-c_\rho f_1
  =A(\rho-c_\rho f_1)A^\dagger+\Delta\succeq0.
  \label{eq:conversion-free-fraction}
\end{equation}
Thus, $\sigma$ has a free-state fraction of at least $c_\rho$, which implies
$c_\sigma\ge c_\rho$.
\end{proof}

We emphasize that the grand sum and the maximal free fraction impose
complementary restrictions on state conversion. The filter preserves the
population in the textureless state, whereas preparation can only increase it.
In contrast, the maximal free fraction $c_\rho$ is the largest weight of $f_1$
in a convex decomposition of $\rho$; the filter preserves such a component,
which stays present after adding the positive semidefinite remainder $\Delta$.
In terms of the geometric texture and texture weight measures, these necessary
conditions read
\begin{equation}
  \mathcal T_g(\sigma)\le\mathcal T_g(\rho),
  \qquad
  \mathcal T_w(\sigma)\le\mathcal T_w(\rho).
  \label{eq:general-two-measures}
\end{equation}
For a single qubit, these two inequalities are sufficient, as shown in
\Eref{eq:qubit-two-measures}. However, in dimensions greater than two, they are
necessary but are not enough in general, as we illustrate with examples in
\Sref{sec:consequences}.

Let us examine the positivity condition $\Delta\succeq0$ in terms of the blocks
of $\Delta$. The state $\omega$ prepared after the second outcome has to
provide enough population in order to support the coherences that the filtered
source misses. Let us write $\alpha=[\Sigma(\sigma)-\Sigma(\rho)]/D$,
$z=y-Tx$, and $H=S-TRT^\dagger$, which correspond, respectively, to the missing
population in the textureless state, the missing coherences with $\ket{f_1}$,
and the remainder block on the maximally textured subspace.
For $\alpha>0$, the condition $\Delta\succeq0$ means that the expectation value
of $\Delta$ on every vector $(a,v)\in\C\oplus\C^{D-1}$, namely
$\alpha|a|^2+2\operatorname{Re}(a^*z^\dagger v)+v^\dagger Hv$, is nonnegative.
Completing the square in $a$ by adding and subtracting $|z^\dagger v|^2/\alpha$,
one obtains
\begin{equation}
  \alpha\left|a+\frac{z^\dagger v}{\alpha}\right|^2
  +v^\dagger\left(H-\frac{zz^\dagger}{\alpha}\right)v.
\end{equation}
For each $v$, setting $a=-z^\dagger v/\alpha$ makes the first term vanish.
Hence, $\Delta\succeq0$ holds if and only if $H-zz^\dagger/\alpha\succeq0$,
that is,
\begin{equation}
  S-TRT^\dagger
  -\frac{D(y-Tx)(y-Tx)^\dagger}
        {\Sigma(\sigma)-\Sigma(\rho)}\succeq0.
  \label{eq:schur-general}
\end{equation}
This Schur-complement condition requires the remainder $H$ on the maximally
textured subspace to supply the positive contribution $zz^\dagger/\alpha$ needed to
support the missing coherences. If the filter already supplies all
these coherences ($z=0$), only $H\succeq0$ is required.

At equal grand sum, preparing $\omega$ adds no population along the
textureless subspace and therefore cannot supply any coherence with it.
When $\alpha=0$, the expectation value above reduces to
$2\operatorname{Re}(a^*z^\dagger v)+v^\dagger Hv$.
If $z\ne0$, we can pick a vector $v$ such that $z^\dagger v\ne0$ and then
adjust the phase and magnitude of $a$ to drive this quantity negative.
For $\Delta$ to be positive semidefinite, we therefore need $z=0$; in other
words, the filter alone has to produce all coherences with the textureless
state. All that remains is $H\succeq0$, which gives
\begin{equation}
  y=Tx,
  \qquad
  S-TRT^\dagger\succeq0.
  \label{eq:qeqp}
\end{equation}
In this situation, $\omega$ can be taken to be a maximally textured state,
which gives a free operation that conserves the grand sum on every input state.

\begin{corollary}[Conversion at constant grand sum]
\label{cor:constant-grand-sum-conversion}
If $\rho\longrightarrow\sigma$ and $\Sigma(\sigma)=\Sigma(\rho)$,
the conversion can be realized by a QST free operation that preserves
the grand sum of every state. If $R\succ0$, every QST free operation
realizing the conversion has this property.
\end{corollary}
\begin{proof}
Let us start with a channel given by Theorem~\ref{thm:exact-convertibility}.
When $p_{\mathrm{prep}}>0$, we have
$0=p_{\mathrm{prep}}\mel{f_1}{\omega}{f_1}$, so $\mel{f_1}{\omega}{f_1}=0$.
Since $\omega\succeq0$, its support must lie inside the maximally textured
subspace. When $p_{\mathrm{prep}}=0$, we are free to pick any $\omega$ with
support in that subspace without affecting the output on $\rho$. On any input
$X$, the filter does not change the population in $f_1$, and preparing $\omega$
contributes nothing to it either, so $\Tr[f_1\Lambda(X)]=\Tr(f_1X)$.

Now suppose $R\succ0$. For any QST free operation that converts $\rho$ into
$\sigma$, \Eref{eq:grand-sum-increment} forces $\Tr[\Phi^\dagger(f_1)R]=0$.
Because $R$ is strictly positive, there exists $r>0$ with
$R\succeq r\id_{D-1}$. Together with $\Phi^\dagger(f_1)\succeq0$, this leads
to $0\ge r\Tr[\Phi^\dagger(f_1)]\ge0$, and hence $\Phi^\dagger(f_1)=0$. By
Corollary~\ref{cor:grand-sum-preservation}, the channel preserves the grand sum
of every state.
\end{proof}

Notice that keeping the grand sum constant still leaves room for the
preparation branch to modify the state inside the maximally textured subspace.
When $R\succ0$, every free operation achieving the conversion has to preserve
the grand sum on all inputs; if $R$ is rank deficient, however, there may be
other channels realizing $\rho\longrightarrow\sigma$ that increase the grand
sum on inputs occupying directions in the maximally textured subspace absent
from $\rho$.

In several important cases, we do not even need to run the SDP because we can
write down the filter directly. Let us begin with a maximally textured source.
A measurement distinguishes this source from the textureless state with
certainty, so the second outcome occurs with probability one and we can
prepare any target.

\begin{corollary}[Maximally textured states]
\label{cor:universal-sources}
A state $\rho$ can reach every target under QST free operations if and only
if it is maximally textured, meaning that its support
lies entirely in the maximally textured subspace. Equivalently,
\begin{equation}
  \Sigma(\rho)=D\mel{f_1}{\rho}{f_1}=0.
  \label{eq:universal-p0}
\end{equation}
\end{corollary}
\begin{proof}
If $\rho$ is maximally textured, then
$\rho=f_1^{\perp}\rho f_1^{\perp}$.
We use $T=0$, which satisfies \Eref{eq:literature-contraction},
and $\omega=\sigma$ for an arbitrary target $\sigma$.
\Eref{eq:conversion-residual} in
Theorem~\ref{thm:exact-convertibility} then gives
$\Phi(Z)=\Tr(Z)\sigma$.
Theorem~\ref{thm:free-normal-form-literature} yields the QST free channel
in \Eref{eq:replacement-free-channel}, with $A=f_1$.
The filter output vanishes on $\rho$ because
$A\rho A^\dagger=f_1\rho f_1=0$, while the residual input
$R=f_1^{\perp}\rho f_1^{\perp}$ has trace one.
Thus $\Phi(R)=\sigma$ and $\Lambda(\rho)=\sigma$.
For the converse, let us assume that the source $\rho$ can reach any target
state, and in particular a maximally textured target $\sigma$. Such a target
has $\Sigma(\sigma)=0$, so \Eref{eq:p-monotone} immediately implies
$\Sigma(\rho)=0$. Because $\rho\succeq0$, we can write
$\mel{f_1}{\rho}{f_1}=\norm{\rho^{1/2}\ket{f_1}}^2$; thus, $\Sigma(\rho)=0$
is equivalent to $\rho\ket{f_1}=0$, that is, the support of $\rho$ lies
entirely in the maximally textured subspace.
\end{proof}

Consequently, the maximally textured states form a single maximal equivalence
class under QST free operations, in the sense that any state in this class can
reach every state, and in particular every other maximally textured state. For that reason, every texture monotone attains its global maximum on
all maximally textured states~\cite{deLeonReachability,MaFourier2026}.

In the case of a pure source state, the grand-sum inequality alone decides
convertibility to any target $\sigma$, independent of the rank or spectrum of
$\sigma$. Maximally textured pure states are already covered by
Corollary~\ref{cor:universal-sources}. When the source state has nonzero
overlap with the textureless state, it is enough to construct a quantum filter
that maps the source vector to an unnormalized vector whose outer product is
bounded above by $\sigma$, while preparation supplies the positive semidefinite
remainder. In this way, we recover the deterministic case of the optimal
conversion probability for pure sources obtained in Ref.~\cite{Lin2026}.

\begin{corollary}[Pure source and arbitrary target]
\label{cor:pure-to-arbitrary}
For a pure source $\rho=\ket{\psi}\bra{\psi}$ and any target
density operator $\sigma$,
\begin{equation}
  \rho\longrightarrow\sigma
  \quad\Longleftrightarrow\quad
  \Sigma(\sigma)\ge\Sigma(\rho).
  \label{eq:pure-input}
\end{equation}
\end{corollary}
\begin{proof}
The necessity of the inequality follows from \Eref{eq:p-monotone}.
In order to prove sufficiency, let us assume $\Sigma(\sigma)\ge\Sigma(\rho)$.
If $\Sigma(\rho)=0$, the conversion is given by
Corollary~\ref{cor:universal-sources}. If $\Sigma(\rho)=D$, then $\rho=f_1$,
and the inequality implies $\Sigma(\sigma)=D$, so $\sigma=f_1$ as well.
We are therefore left with the case $0<\Sigma(\rho)<D$, in which
$\Sigma(\sigma)>0$. By choosing the global phase of $\ket{\psi}$ properly, we
write $\ket{\psi}=\sqrt{\frac{\Sigma(\rho)}{D}}\,\ket{f_1}+\ket{u}$, where
$\ket{u}$ belongs to the maximally textured subspace and satisfies
$\norm{u}^2=1-\frac{\Sigma(\rho)}{D}$.

We seek a filtered vector with the same amplitude
$\sqrt{\frac{\Sigma(\rho)}{D}}$ in the textureless subspace and an outer
product bounded above by $\sigma$.
We use
\begin{equation}
  \ket{a}=\frac{\sqrt{D\Sigma(\rho)}}{\Sigma(\sigma)}\,
    \sigma\ket{f_1}.
\end{equation}
It satisfies $\braket{f_1}{a}=\sqrt{\frac{\Sigma(\rho)}{D}}$.
To check the bound on the outer product, we apply Cauchy--Schwarz to
$\sigma^{1/2}\ket{z}$ and $\sigma^{1/2}\ket{f_1}$. This gives
\begin{equation}
  \abs{\mel{z}{\sigma}{f_1}}^2
  \le \frac{\Sigma(\sigma)}{D}\mel{z}{\sigma}{z}
\end{equation}
for every vector $z$. Equivalently,
$\sigma f_1\sigma\preceq \frac{\Sigma(\sigma)}{D}\sigma$. Therefore
\begin{equation}
  \ket{a}\bra{a}
  =\frac{D\Sigma(\rho)}{\Sigma(\sigma)^2}\sigma f_1\sigma
  \preceq\frac{\Sigma(\rho)}{\Sigma(\sigma)}\sigma
  \preceq\sigma.
\end{equation}
Taking traces shows that $\norm{a}^2\le1$.
Since its textureless-subspace component has squared norm $\frac{\Sigma(\rho)}{D}$, the remaining
component obeys $\norm{f_1^{\perp}a}^2\le1-\frac{\Sigma(\rho)}{D}=\norm{u}^2$.

To map $u$ to $f_1^{\perp}a$, we use the contraction
\begin{equation}
  T=\frac{\ket{f_1^{\perp}a}\bra{u}}{1-\frac{\Sigma(\rho)}{D}}.
\end{equation}
Its squared norm satisfies
$\norm{f_1^{\perp}a}^2/(1-\frac{\Sigma(\rho)}{D})\le1$, and the corresponding
filter gives $A\ket{\psi}=\ket{a}$.
The remainder $\Delta=\sigma-\ket{a}\bra{a}$ is positive semidefinite.
If $p_{\mathrm{prep}}=\Tr\Delta>0$, we choose $\omega=\Delta/p_{\mathrm{prep}}$,
whereas if $p_{\mathrm{prep}}=0$, any density operator $\omega$ can be used.
In both cases, the residual map of Theorem~\ref{thm:exact-convertibility}
provides the remainder $\Delta$, which proves the conversion.
\end{proof}

Immediately, it follows that two pure states are interconvertible if and only
if their grand sums are equal, and the conversion can be realized by a unitary
QST free operation. Indeed, we can fix their overall phases in such a way that
their amplitudes along $\ket{f_1}$ coincide; their components in the maximally
textured subspace then have the same norm. A unitary operator on the maximally
textured subspace maps one component into the other while leaving $\ket{f_1}$
unchanged, without any need for residual preparation.

In the complementary case of a pure target and an arbitrary, possibly mixed,
source, state conversion depends on the overlap of the support of the source
with the textureless state. Because the target is pure, every nonzero
contribution to the output has to share the same one-dimensional support. In
turn, this yields an explicit criterion written in terms of the entire support
of the source instead of its population in the textureless state.
\begin{corollary}[Arbitrary source and pure target]
For any density operator $\rho$ and pure target
$\sigma=\ket{\phi}\bra{\phi}$, we use $P_\rho$ for the
orthogonal projector onto $\supp\rho$. Explicitly, if
$\rho=\sum_j\lambda_j\ket{r_j}\bra{r_j}$ is a spectral decomposition
in an orthonormal eigenbasis, then
\begin{equation}
  P_\rho=\sum_{j:\lambda_j>0}\ket{r_j}\bra{r_j}.
\end{equation}
Thus $P_\rho$ projects onto the span of the eigenvectors of $\rho$
with nonzero eigenvalues. The conversion criterion is
\begin{equation}
  \rho\longrightarrow\sigma
  \quad\Longleftrightarrow\quad
  \mel{f_1}{P_\rho}{f_1}\le\frac{\Sigma(\sigma)}{D}.
\end{equation}
\end{corollary}
\begin{proof}
We write $p=\Sigma(\rho)/D$, $q=\Sigma(\sigma)/D$, and
$s=\mel{f_1}{P_\rho}{f_1}$. To prove necessity, note that
Theorem~\ref{thm:exact-convertibility} gives a contraction
$A=\operatorname{diag}(1,T)$ obeying
$0\preceq A\rho A^\dagger\preceq\sigma$. Because $\sigma=\ket{\phi}\bra{\phi}$
is rank one, the image of $\supp\rho$ under $A$ has to fall inside the span of
$\ket{\phi}$. When $q>0$, let us fix the phase of $\ket{\phi}$ so that
$\braket{f_1}{\phi}=\sqrt q$. Combining this with $\bra{f_1}A=\bra{f_1}$, we
find
\begin{equation}
  AP_\rho=\frac{\ket{\phi}\bra{f_1}P_\rho}{\sqrt q},
  \qquad
  \norm{AP_\rho}^2=\frac{s}{q}\le1,
\end{equation}
where the norm is the operator norm, and therefore $s\le q$.
When $q=0$, \Eref{eq:p-monotone} implies $p=0$, which gives $\rho\ket{f_1}=0$
and hence $s=0$ as well.
Conversely, suppose $s\le q$. If $q=0$, then $s=0$ too, so $\rho$ is maximally
textured and Corollary~\ref{cor:universal-sources} already gives the
conversion. If $q=1$, the target state is $f_1$, which can always be reached by means of
the complete reset $X\mapsto\Tr(X)f_1$.
It remains to treat $0<q<1$, where we also have $s<1$. With the same phase
convention for $\ket{\phi}$, let us define
\begin{equation}
  \ket{v}=f_1^\perp\ket{\phi},
  \qquad
  \ket{u}=f_1^\perp P_\rho\ket{f_1},
\end{equation}
and consider the operator on the maximally textured subspace
\begin{equation}
  T=\frac{\ket{v}\bra{u}}{\sqrt q(1-s)}.
\end{equation}
Since $\norm{u}^2=s(1-s)$ and $\norm{v}^2=1-q$, we find
\begin{equation}
  \norm{T}^2=\frac{(1-q)s}{q(1-s)}\le1,
\end{equation}
so $T$ is indeed a contraction.
In addition, $\bra{u}P_\rho=(1-s)\bra{f_1}P_\rho$, which implies that the
corresponding filter $A$ obeys the same identity for $AP_\rho$ derived in the
necessity proof. From this, we obtain
\begin{equation}
  A\rho A^\dagger=\frac{p}{q}\sigma,
  \qquad
  \Delta=\left(1-\frac{p}{q}\right)\sigma\succeq0,
\end{equation}
since $p\le s\le q$ follows from $\rho\preceq P_\rho$.
Taking $\omega=\sigma$ in Theorem~\ref{thm:exact-convertibility} gives this
remainder and completes the proof.
\end{proof}
It is worth noting that the support of the source state alone determines which
pure targets are reachable; the nonzero eigenvalues of $\rho$ do not introduce
any additional constraint. For a pure source state, $P_\rho=\rho$, and we
recover the grand-sum inequality of Corollary~\ref{cor:pure-to-arbitrary}.
\endgraf
Whenever the support of the source contains $\ket{f_1}$, the only pure target
that can be reached is the textureless state $f_1$. This is the case, in
particular, for every full-rank source state. For instance, the maximally mixed
state $\rho=\id_D/D$ has $P_\rho=\id_D$, so the criterion imposes
$\Sigma(\sigma)/D\ge1$, which forces $\sigma=f_1$.
On the other hand, mixed states that are not full rank can reach textured pure
targets provided that $\ket{f_1}$ does not belong to their support.
Equivalently, we can write the criterion as an upper bound on the geometric
texture of a reachable pure target,
\begin{equation}
  \mathcal T_g(\sigma)\le
  \mel{f_1}{(\id_D-P_\rho)}{f_1}.
\end{equation}

In this way, the criterion of a positive remainder gives both a test for state
convertibility and an explicit protocol whenever the conversion is allowed. We
have obtained analytical filters for maximally textured sources and for
conversions in which at least one state is pure. When both the source and
target states are mixed, however, the two scalar conditions in
\Eref{eq:general-two-measures} are not enough in general, and one has to solve
a matricial feasibility problem. In \Sref{sec:consequences}, we illustrate our
results with concrete conversion families and study the qutrit case, where
coherences can prevent conversion even when both scalar conditions hold.

\section{Illustrative conversions and the qutrit transition}
\label{sec:consequences}

In this section, we apply the normal form of QST free operations and the
criterion of a positive remainder to concrete families of state conversions.
We first discuss conversions realized by measurement and preparation, and we
then study coherent qutrit states in order to see how conversion can fail even
when both geometric texture and texture weight obey the necessary inequalities.

\begin{figure}[t]
  \centering
  \includegraphics[width=\columnwidth]{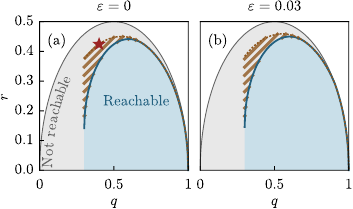}
  \caption{Qutrit reachability beyond geometric texture and texture weight.
  For source $\rho$ in \Eref{eq:counterexample}, targets are
  $\sigma_\varepsilon(q,r)$ [\Eref{eq:numerical-target-family}] with
  (a) $\varepsilon=0$ and (b) $\varepsilon=0.03$.
  Blue regions are SDP-reachable; gray regions are unreachable.
  Brown hatching marks unreachable targets satisfying both monotone constraints
  [\Eref{eq:numerical-scalar-region}], whose upper bound is dashed brown.
  The gray curve is $r=\sqrt{q(1-q)}$; the red star marks the target $\sigma$ in \Eref{eq:counterexample}.
  All targets in (b) are full rank.}
  \label{fig:qutrit-reachability}
\end{figure}

\subsection{Conversion by measurement and preparation}

When neither the source nor the target has coherence between the textureless
state and the maximally textured subspace, measurement and preparation are
enough to perform the conversion.
Let us first consider the case in which the target has a larger grand sum, so
population has to be transferred into the textureless state.
For $x=y=0$ in \Eref{eq:block-states}, we choose $T=0$.
In this case, the filter is simply $A=f_1$, which keeps only the component in
the textureless state, whereas the residual map $\Phi$ supplies the whole block
of the output state on the maximally textured subspace. The remainder in
\Eref{eq:Delta} reduces to
\begin{equation}
  \Delta=\frac{\Sigma(\sigma)-\Sigma(\rho)}{D}f_1+S,
\end{equation}
where $S$ is the block of $\sigma$ on the maximally textured subspace in
\Eref{eq:block-states}.
Because $S\succeq0$, checking the positivity of $\Delta$ reduces to checking
its entry in the textureless state. More generally, in every dimension,
\begin{equation}
  x=y=0:\qquad
  \rho\longrightarrow\sigma\quad\Longleftrightarrow\quad \Sigma(\sigma)\ge\Sigma(\rho).
  \label{eq:block-diag-family}
\end{equation}
For $\Sigma(\rho)<D$, the two measurement outcomes have probabilities $\Sigma(\rho)/D$
and $1-\Sigma(\rho)/D$: after the first, we retain $f_1$; after the second, we prepare
\begin{equation}
  \omega=\frac{[\Sigma(\sigma)-\Sigma(\rho)]f_1/D+S}{1-\Sigma(\rho)/D}.
  \label{eq:block-diag-preparation}
\end{equation}
This is the measurement-and-preparation channel of
\Eref{eq:replacement-free-channel}. Its residual map supplies both
the target block on the maximally textured subspace and the increase in
textureless population.

At equal grand sum, preparation can instead change the state within
the maximally textured subspace without transferring population between
subspaces.
In particular, for
\begin{equation}
  \rho=\frac{\Sigma(\rho)}{D}f_1+\left(1-\frac{\Sigma(\rho)}{D}\right)\tau_\perp,
  \qquad f_1^\perp\tau_\perp f_1^\perp=\tau_\perp,
  \label{eq:rho-p}
\end{equation}
at $\Sigma(\sigma)=\Sigma(\rho)<D$, the preparation can replace
$\tau_\perp$ by any $\omega_\perp$. Block-diagonal states with the same grand sum are therefore mutually
convertible even when their spectra differ. Such conversions need not
be unitary.

\subsection{A coherent filter-and-preparation protocol}

The filter and the residual map can supply coherences along mutually
orthogonal directions in the maximally textured subspace.
In the adapted qutrit basis
$\{\ket{f_1},\ket{e_1},\ket{e_2}\}$, let us use the source family
\begin{equation}
  \rho=\begin{pmatrix}
    a&z^*&0\\z&b&0\\0&0&1-a-b
  \end{pmatrix},
  \label{eq:coherent-example}
\end{equation}
where $a,b>0$, $a+b<1$, and $0<|z|^2<ab$. These conditions ensure
that $\rho$ is full rank and has coherence between $f_1$ and $e_1$.
For $0\le\eta\le1$, we use
$T_\eta=\operatorname{diag}(\sqrt\eta,0)$ and prepare
$\omega=\dyad{+}$, where $\ket{+}=(\ket{f_1}+\ket{e_2})/\sqrt2$.
The filter retains population $a+\eta b$, leaving preparation probability
$p_{\mathrm{prep},\eta}=1-a-\eta b>0$. The resulting targets are
\begin{equation}
  \sigma_\eta=\begin{pmatrix}
    a+p_{\mathrm{prep},\eta}/2&\sqrt\eta\,z^*&p_{\mathrm{prep},\eta}/2\\
    \sqrt\eta\,z&\eta b&0\\
    p_{\mathrm{prep},\eta}/2&0&p_{\mathrm{prep},\eta}/2
  \end{pmatrix}.
  \label{eq:coherent-target-family}
\end{equation}
Indeed, with $A_\eta=\operatorname{diag}(1,\sqrt\eta,0)$,
\begin{equation}
  \Delta_\eta=\sigma_\eta-A_\eta\rho A_\eta^\dagger
  =p_{\mathrm{prep},\eta}\dyad{+}\succeq0,
  \label{eq:coherent-completion}
\end{equation}
so \Eref{eq:filter-reprepare-channel} realizes every conversion
$\rho\longrightarrow\sigma_\eta$.

The two contributions control distinct target coherences throughout this
family. The filter multiplies the $f_1$--$e_1$ coherence by $\sqrt\eta$,
while preparation supplies the entire $f_1$--$e_2$ coherence and raises
the population in the textureless subspace from $a$ to $a+p_{\mathrm{prep},\eta}/2$.

Unitary QST free operations are the case in which no preparation
is needed. For $T=V\in U(D-1)$, the defect vanishes, and
$U_f=\operatorname{diag}(1,V)$ connects states satisfying
\begin{equation}
  \Sigma(\sigma)=\Sigma(\rho),\qquad y=Vx,\qquad S=VRV^\dagger.
  \label{eq:unitary-orbit}
\end{equation}
The inverse unitary realizes the reverse conversion.

\subsection{Explicit qutrit conversion tests}

In the case of qutrits ($D=3$), the matricial conditions for state conversion
reduce to trace and determinant inequalities. Because the maximally textured
subspace is two dimensional, $T$, $R$, and $S$ are $2\times2$ matrices. When
$\Sigma(\sigma)>\Sigma(\rho)$, the Schur-complement condition in
\Eref{eq:schur-general} can be written in terms of the $2\times2$ matrix
\begin{equation}
  Q_T:=S-TRT^\dagger
  -\frac{3(y-Tx)(y-Tx)^\dagger}{\Sigma(\sigma)-\Sigma(\rho)}.
  \label{eq:QT}
\end{equation}
Then, conversion is possible if and only if there exists a matrix $T$
satisfying
\begin{equation}
  T^\dagger T\preceq\id_2,
  \qquad Q_T\succeq0.
  \label{eq:qutrit-two-PSD}
\end{equation}
Since a Hermitian $2\times2$ matrix is positive semidefinite if and only if
both its trace and its determinant are nonnegative, \Eref{eq:qutrit-two-PSD}
is equivalent to the four scalar inequalities
\begin{subequations}
\begin{align}
  \Tr(\id_2-T^\dagger T)&\ge0,\\
  \det(\id_2-T^\dagger T)&\ge0,\\
  \Tr Q_T&\ge0,\\
  \det Q_T&\ge0.
\end{align}
\label{eq:qutrit-scalars}
\end{subequations}
When the source and target have the same grand sum, \Eref{eq:qeqp} imposes
$y=Tx$ and $S-TRT^\dagger\succeq0$, along with the contraction condition
$T^\dagger T\preceq\id_2$. In that case, the same trace and determinant tests
apply to the block $S-TRT^\dagger$ on the maximally textured subspace. We note
that neither case assumes $R$ or $S$ to be invertible.

These conditions keep information about the conversion that geometric texture
and texture weight do not capture in general. Unlike the single-qubit case, $T$ is now a
$2\times2$ matrix acting on two independent directions, and its matrix elements
have to leave a positive semidefinite remainder without increasing the norm of
any vector. As we show in the next example, even a single diagonal element of
$Q_T$ can force an entry of $T$ that amplifies a vector, which rules out the
conversion even when both texture measures are nonincreasing.

\subsection{An obstruction beyond scalar monotones}
Let us consider the following pair of qutrit states, written in the adapted
basis $\{\ket{f_1},\ket{e_1},\ket{e_2}\}$,
\begin{equation}
  \rho=
  \begin{pmatrix}
    0.3 & \sqrt{0.02} & 0\\
    \sqrt{0.02} & 0.1 & 0\\
    0&0&0.6
  \end{pmatrix},
  \quad
  \sigma=
  \begin{pmatrix}
    0.4 & \sqrt{0.18} & 0\\
    \sqrt{0.18} & 0.6 & 0\\
    0&0&0
  \end{pmatrix}.
  \label{eq:counterexample}
\end{equation}
For these states, we obtain
\begin{equation}
  \Sigma(\sigma)>\Sigma(\rho),
  \qquad
  c_\rho=c_\sigma=0.1,
  \label{eq:counter-scalars}
\end{equation}
or, expressed in terms of geometric texture and texture weight,
\begin{equation}
  \begin{aligned}
    \mathcal T_g(\rho)&=0.7, &\mathcal T_g(\sigma)&=0.6,\\
    \mathcal T_w(\rho)&=0.9, &\mathcal T_w(\sigma)&=0.9.
  \end{aligned}
  \label{eq:counter-measures}
\end{equation}
Hence, both scalar monotonicity conditions are satisfied. However, if we write
the matrix elements of the filter as $T=(t_{jk})_{j,k=1}^2$ and substitute the
blocks of $\rho$ and $\sigma$ into \Eref{eq:QT}, we find for the upper-left
entry of $Q_T$
\begin{equation}
  \begin{aligned}
    (Q_T)_{11}
    &=\frac35-\frac1{10}|t_{11}|^2-\frac35|t_{12}|^2
      -\frac15|3-t_{11}|^2\\
    &=-\frac{3}{10}|t_{11}-2|^2-\frac35|t_{12}|^2.
  \end{aligned}
  \label{eq:counter-diagonal-remainder}
\end{equation}
Because $Q_T\succeq0$ requires every diagonal entry of $Q_T$ to be nonnegative,
we are forced to take
\begin{equation}
  t_{11}=2,\qquad t_{12}=0.
  \label{eq:counter-filter-constraint}
\end{equation}
On the other hand, any contraction has to obey $\|T\ket{e_1}\|\le1$, whereas
$t_{11}=2$ implies $\|T\ket{e_1}\|\ge2$. Immediately, it follows that
\begin{equation}
  \rho\not\longrightarrow\sigma.
  \label{eq:counter-no}
\end{equation}
In physical terms, reaching $\sigma$ would require the filter to amplify the
component along $\ket{e_1}$ by a factor of two, which is impossible for a
quantum filter. Moreover, the reverse conversion $\sigma\longrightarrow\rho$ is
ruled out by the monotonicity of geometric texture, so $\rho$ and $\sigma$ are
incomparable under QST free operations. Therefore, geometric texture and
texture weight are not enough to decide state convertibility for qutrits,
although they form a complete pair of monotones for a single qubit.

We stress that this obstruction occurs across a finite region of target states,
including full-rank targets. In order to illustrate this, we keep the source
state $\rho$ of \Eref{eq:counterexample} and consider the family of qutrit
targets
\begin{equation}
  \sigma_\varepsilon(q,r)=(1-\varepsilon)
  \begin{pmatrix}
    q&r&0\\r&1-q&0\\0&0&0
  \end{pmatrix}+\frac{\varepsilon}{3}\id_3,
  \label{eq:numerical-target-family}
\end{equation}
with $0\le q\le1$, $0\le r\le\sqrt{q(1-q)}$, and $0\le\varepsilon<1$. If we set
$q_\varepsilon=(1-\varepsilon)q+\varepsilon/3$ and
$s_\varepsilon=(1-\varepsilon)(1-q)+\varepsilon/3$, the two scalar constraints
take the form
\begin{equation}
  q_\varepsilon\ge0.3,\qquad
  (1-\varepsilon)^2r^2\le(q_\varepsilon-0.1)s_\varepsilon.
  \label{eq:numerical-scalar-region}
\end{equation}
For fixed values of $q$ and $\varepsilon$, we maximize $r$ subject to the
feasibility conditions in \Eref{eq:conversion-sdp} and the physical range
above. Because $\sigma_\varepsilon(q,r)$ depends affinely on $r$, this
optimization is still an SDP. As shown in \Fref{fig:qutrit-reachability}, the
resulting boundary separates the reachable targets from a whole region of
states that satisfy both scalar inequalities but cannot be reached under the
exact criterion. In particular, this unreachable region remains present at
$\varepsilon=0.03$, where every target state is full rank with minimum
eigenvalue at least $0.01$.

Finally, the same counterexample extends to every dimension $D\ge3$. We simply
embed the qutrit states of \Eref{eq:counterexample} into the subspace spanned
by $\{\ket{f_1},\ket{e_1},\ket{e_2}\}$ and assign zero support to the remaining
basis vectors. In \Eref{eq:schur-general}, these zero blocks leave the same
diagonal entry unchanged, so adding extra dimensions cannot remove the
contradiction. In the case of a single qubit ($D=2$), however, the maximally
textured subspace is one dimensional and $T$ reduces to a single complex
scalar, which is why the single-qubit conversion problem collapses to two
scalar inequalities~\cite{Lin2026,deLeon2026}. Thus, going from qubits to
qutrits already brings in the matricial structure that governs state conversion
in dimensions greater than two.

\section{Conclusions}
\label{sec:conclusions}

In this article, we demonstrate that exact state conversion under QST free operations can be decided in
arbitrary finite dimensions by asking whether a quantum filter leaves a physically
allowed remainder. We obtain this criterion by first deriving a general
characterization of QST free operations. These separate into a
quantum filter that represents one measurement outcome and a residual
completely positive map that acts only on the input block in the maximally
textured subspace. Such a filter preserves the unique QST free state and
can rotate and attenuate the components in the maximally textured subspace.
For a given input state, the filter determines the probability of the
complementary outcome, which fixes the trace that the residual map has to supply.
State conversion is thus possible when some such filter leaves a 
positive semidefinite remainder (an unnormalized state or the zero operator), 
that is, the target minus the unnormalized filtered source.

The state conversion criterion gives both a computational test and a physical protocol
for every allowed conversion. The filter can be determined by solving a semidefinite
program that includes matricial constraints.
Once the filter is found, a measurement with two outcomes followed by
conditional preparation realizes the conversion.
One outcome retains the filtered state, and the other prepares a fixed state
that supplies the remainder obtained by subtracting the unnormalized filtered
source state from the target.
This construction confirms that maximally textured states can reach, that is, 
can be converted to, any target state. Their maximal conversion power comes
from the fact that they are perfectly distinguishable from the textureless
state, hence allowing any target state to be prepared while leaving the
textureless state unchanged.

On the one hand, as a byproduct of this work, we give new insight into state
conversion for a single qubit by reformulating it in terms of the grand sum
and the maximal free fraction. On the other, our general criterion establishes,
as we illustrate with examples, that these two quantities are
not enough to decide conversion in dimensions greater than two.
The grand sum captures the population in the textureless state.
The maximal free fraction is the largest weight that the textureless state can
have in a convex decomposition of the density matrix. 
Therefore, we argue they stand on equal footing in the way they
capture the resource of texture, and we prove they are
complementary; neither quantity determines the other.
Our reformulation of the single-qubit state conversion criterion in terms 
of those two quantities illustrates their complementarity.
However, beyond a single qubit, they are unable to decide convertibility in general.

Some interesting questions now arise for future work.
To begin, it must be stressed that the results of this article are applicable
to any quantum resource theory with a single pure state as its only free
state, see, e.g., Ref.~\cite{Greenwood2026}. Phase transitions would be
interesting to investigate through the lens of state convertibility,
including studying whether a preorder of states exists
and how it changes at a critical point~\cite{Patra2026,Celeri2026,Casagrande2026,Dittberner2026,XuTexture2026}.
The quantum Mpemba effect~\cite{Summer2026} is another quantum phenomenon that could be studied
in QST using the criterion of convertibility. On the mathematical front,
deriving conditions for state convertibility in terms of the source and
target states would be valuable. Such conditions would be distinct from our
criterion that yields a free operation achieving the conversion. Finally,
studying Lindbladian QST free dynamics could build on our results and would
represent an interesting avenue.

\begin{acknowledgments} 
The author thanks Miguel Gonzalez for helpful discussions that made him 
recognize the importance of the texture weight measure.  
Fernando Parisio, Pedro C. Azado, and Alejandro Fonseca are thanked 
for discussions and comments on the manuscript. Special
thanks to Carlos Pineda for his always-tough feedback on improving
the readability of the manuscript.
Support from SECIHTI for graduate studies and through grant CBF-2025-I-1548,
and from UNAM-PAPIIT through grant IG101324, is acknowledged.
The author discloses using OpenAI's Codex (GPT-6-Sol and GPT-6.1-Sol) 
for manuscript editing, literature
assessment, and exploration and cross-checking of derivations, all based
on research ideas developed entirely by the author. He independently
verified all material incorporated into the manuscript and takes full
responsibility for this work.
\end{acknowledgments} 

\appendix
\section{Proof of Theorem~\ref{thm:free-normal-form-literature}}
\label{app:normal-form}

\begin{proof}
We start with a CPTP map $\Lambda$ that fixes $f_1$ and a Kraus
representation $\Lambda(X)=\sum_{j=1}^{r}K_jXK_j^\dagger$. We seek a new
representation in which one operator leaves $\ket{f_1}$ unchanged and
all the others annihilate it. These remaining operators will define the
residual map $\Phi$.

The fixed-point condition gives $\Tr[f_1^{\perp}\Lambda(f_1)]=0$. In the Kraus
representation this trace is $\sum_j\norm{f_1^{\perp}K_j\ket{f_1}}^2$, a sum of
nonnegative terms.
Each term must therefore vanish, so $K_j\ket{f_1}=a_j\ket{f_1}$ for
every $j$. In the adapted decomposition, this gives the triangular structure
\begin{equation}
  K_j=
  \begin{pmatrix}
    a_j&b_j\\
    0&B_j
  \end{pmatrix},
\end{equation}
where $a_j\in\C$, $b_j\in M_{1,D-1}(\C)$, and $B_j\in M_{D-1}(\C)$
\cite{NarasimhacharGour2015,Greenwood2026}. The upper-left and upper-right
blocks of the trace-preservation condition
$\sum_jK_j^\dagger K_j=\id_D$ give $\sum_j\abs{a_j}^2=1$ and
$\sum_j a_j^*b_j=0$, respectively.

Because $(a_1^*,\ldots,a_r^*)$ is normalized, we can choose it as the
first row of a unitary matrix $U$ and complete the remaining rows to an
orthonormal basis. Using the unitary freedom of Kraus representations,
we obtain $\widetilde K_\mu=\sum_jU_{\mu j}K_j$, with $\mu=0,\ldots,r-1$.

The first operator is $A:=\widetilde K_0=\sum_j a_j^*K_j$. Its upper-left
entry is $\sum_j\abs{a_j}^2=1$, and its upper-right block is
$\sum_j a_j^*b_j=0$. Thus, defining $T=\sum_j a_j^*B_j$, we obtain
\begin{equation}
  A=\begin{pmatrix}1&0\\0&T\end{pmatrix}.
\end{equation}

For every $\mu>0$, orthogonality of the rows of $U$ to its first row implies
$\sum_jU_{\mu j}a_j=0$, which gives $\widetilde K_\mu\ket{f_1}=0$. As a
consequence, the first column of each remaining Kraus operator vanishes. In the
adapted basis, we can therefore write
\begin{equation}
  \widetilde K_\mu=\begin{pmatrix}0_{D\times1}&C_\mu\end{pmatrix},
  \qquad C_\mu\in M_{D,D-1}(\C),
\end{equation}
where $C_\mu$ consists of the last $D-1$ columns of $\widetilde K_\mu$ and maps
the maximally textured subspace into the system's full Hilbert space.

In this Kraus representation, the trace-preservation condition becomes
\begin{equation}
  \sum_{\mu>0}C_\mu^\dagger C_\mu
  =\id_{D-1}-T^\dagger T\succeq0.
  \label{eq:literature-residual-kraus}
\end{equation}
Taking the expectation value in any vector $v$ in the maximally textured
subspace, this identity reads $\sum_{\mu>0}\norm{C_\mu v}^2=\norm{v}^2-\norm{Tv}^2$. In particular, $T$
cannot increase vector norms, because the remaining Kraus operators account for
the difference between $\norm{v}^2$ and $\norm{Tv}^2$.

Let us now define the map $\Phi(Z)=\sum_{\mu>0}C_\mu ZC_\mu^\dagger$, which is
completely positive because it is written in Kraus form. From
\Eref{eq:literature-residual-kraus}, we obtain
$\Tr\Phi(Z)=\Tr[(\id_{D-1}-T^\dagger T)Z]$. In terms of the Hilbert--Schmidt
adjoint, this relation is \Eref{eq:literature-defect}. Furthermore, each
operator $\widetilde K_\mu$ with $\mu>0$ acts on $X$ only through
$(\id_D-f_1)X(\id_D-f_1)$, so the sum of their contributions equals
$\Phi(f_1^{\perp}Xf_1^{\perp})$. Adding the term coming from the first Kraus
operator $A$, we arrive at \Eref{eq:literature-channel}.

Conversely, if $T$ and $\Phi$ satisfy
\Eref{eq:literature-normal-form}, we can extend Kraus operators
$C_\mu\in M_{D,D-1}(\C)$ for $\Phi$ to the system's full Hilbert space
by adjoining a zero first column:
\begin{equation}
  K_\mu=\begin{pmatrix}0_{D\times1}&C_\mu\end{pmatrix}\in M_D(\C).
\end{equation}
For $\alpha\in\C$ and any vector $v$ in the maximally textured subspace,
this extension acts as
$K_\mu(\alpha\ket{f_1}+v)=C_\mu v$.
Together with $A$, these are Kraus operators for $\Lambda$, proving complete
positivity. By \Eref{eq:literature-defect}, the lower-right block of
their normalization sum is
$T^\dagger T+\sum_\mu C_\mu^\dagger C_\mu=\id_{D-1}$. Its upper-left
entry is one and its off-diagonal blocks vanish, so the full sum equals
$\id_D$. Hence $\Lambda$ preserves trace. Finally, $A\ket{f_1}=\ket{f_1}$
and every other Kraus operator annihilates $\ket{f_1}$, giving
$\Lambda(f_1)=f_1$.
\end{proof}

\bibliography{references.bib}

\end{document}